\documentclass[fleqn,usenatbib]{mnras}

\usepackage{newtxtext,newtxmath}

\usepackage[T1]{fontenc}

\DeclareRobustCommand{\VAN}[3]{#2}
\let\VANthebibliography\thebibliography
\def\thebibliography{\DeclareRobustCommand{\VAN}[3]{##3}\VANthebibliography}

\usepackage{graphicx}	
\usepackage{amsmath}	
\usepackage{bm}
\usepackage{anyfontsize}
\usepackage{xcolor} 
\usepackage{ulem}
\newcommand{\newtext}[1]{\textcolor{black}{#1}}

\title[Minkowski functionals HSC-Y1]{Cosmological constraints using Minkowski functionals from the first year data of the Hyper Suprime-Cam.}

\author[Joaquín Armijo]{
Joaquin Armijo,$^{1,2}$\thanks{E-mail: joaquin.armijo@ipmu.jp}
Gabriela A. Marques,$^{3,4}$
Camila P. Novaes,$^{5,1,2}$ 
Leander Thiele,$^{1,2}$
\newauthor
Jessica A. Cowell,$^{1,2,6}$
Daniela Grandón,$^{7}$
Masato Shirasaki,$^{8,9}$
Jia Liu,$^{1,2}$ 
\\
$^{1}$Kavli Institute for the Physics and Mathematics of the Universe (WPI),\\
The University of Tokyo Institutes for Advanced Study (UTIAS), The University of Tokyo, Chiba 277-8583, Japan \\
$^{2}$Center for Data-Driven Discovery, Kavli IPMU (WPI), UTIAS,
The University of Tokyo, Kashiwa, Chiba 277-8583, Japan\\
$^{3}$Fermi National Accelerator Laboratory, Batavia, IL 60510, USA\\
$^{4}$Kavli Institute for Cosmological Physics, University of Chicago, Chicago, IL 60637, USA\\
$^{5}$Instituto Nacional de Pesquisas Espaciais, Av. dos Astronautas 1758, Jardim da Granja, S\~ao Jos\'e dos Campos, SP, Brazil\\
$^{6}$Department of Physics, University of Oxford, Denys Wilkinson Building, Keble Road, Oxford OX1 3RH, United Kingdom\\
$^{7}$Mathematical Institute, Leiden University, Snellius Gebouw, Niels Bohrweg 1, NL-2333 CA Leiden, The Netherlands\\
$^{8}$National Astronomical Observatory of Japan, National Institutes of Natural Science, Mitaka, Tokyo 181-8588, Japan\\
$^{9}$The Institute of Statistical Mathematics, Tachikawa, Tokyo 190-8562, Japan\\
}
\date{Accepted XXX. Received YYY; in original form ZZZ}

\pubyear{\the\year{2024}}

\begin{document}
\label{firstpage}
\pagerange{\pageref{firstpage}--\pageref{lastpage}}
\maketitle

\begin{abstract}
We use Minkowski functionals to analyse weak lensing convergence maps from the first-year data release of the Subaru Hyper Suprime-Cam (HSC-Y1) survey. Minkowski functionals provide a description of the morphological properties of a field, capturing the non-Gaussian features of the Universe matter-density distribution. Using simulated catalogs that reproduce survey conditions and encode cosmological information, we emulate Minkowski functionals predictions across a range of cosmological parameters to derive the best-fit from the data. By applying multiple scales cuts, we rigorously mitigate systematic effects, including baryonic feedback and intrinsic alignments. From the analysis, combining constraints of the angular power spectrum and Minkowski functionals, we obtain $S_8 \equiv \sigma_8\sqrt{\Omega_{{\rm m}}/0.3} = {0.808}_{-0.046}^{+0.033}$ and $\Omega_{\rm m} = {0.293}_{-0.043}^{+0.157}$. These results represent a $40\%$ improvement on the $S_8$ constraints compared to using power spectrum only. \newtext{Minkowski functionals results are consistent with other two-point, and higher order statistics constraints using the same data, being in agreement with CMB results from the Planck $S_8$ measurements. Our study demonstrates the power of Minkowski functionals beyond two-point statistics to constrain and break the degeneracy between $\Omega_{\rm m}$ and $\sigma_8$.}

\end{abstract}

\begin{keywords}
Cosmology: cosmological parameters, Cosmology - observations
\end{keywords}



\section{Introduction}
\begin{figure*}
    \centering
    \begin{tabular}{cccr}
        \includegraphics[width=0.65\columnwidth]{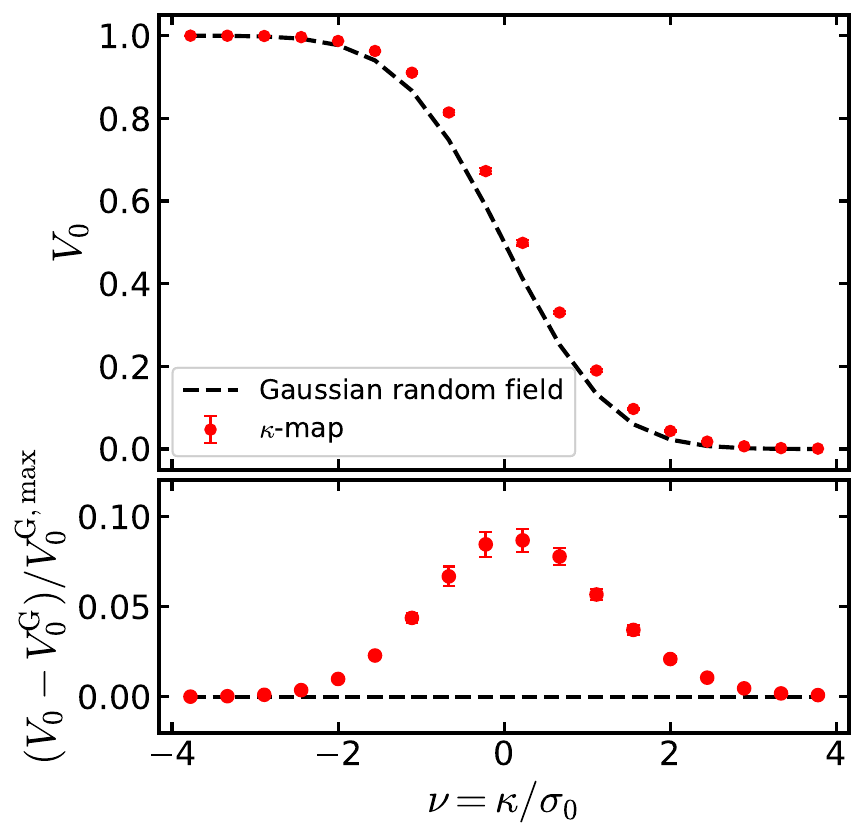} &
        \includegraphics[width=0.65\columnwidth]{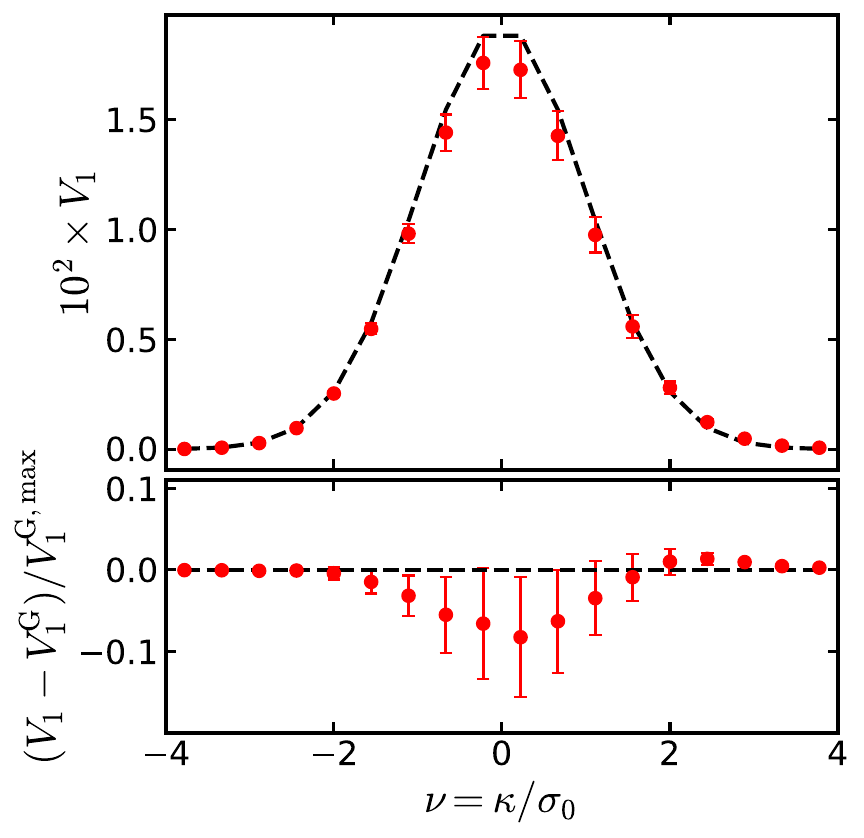} &
        \includegraphics[width=0.65\columnwidth]{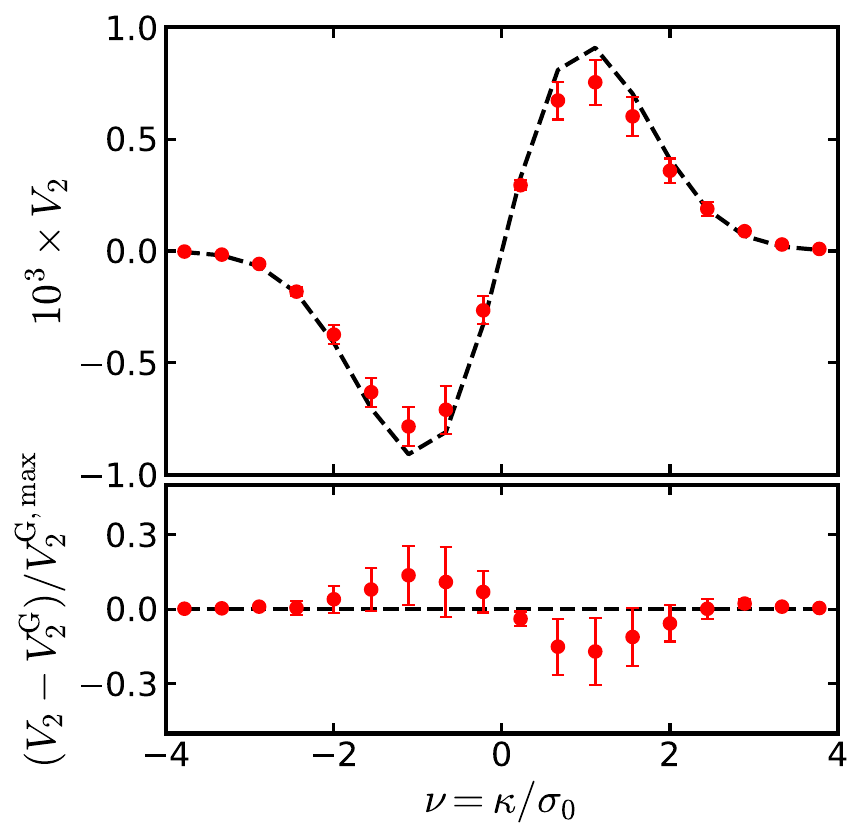} &
    \end{tabular}
\caption{Minkowski Functionals $V_0$ (left), $V_1$ (middle) and $V_2$ (right) as function of $\kappa$ scaled by the standard deviation of the fiducial kappa maps, $\sigma_0$, calculated for the HSC-Y1 data using a $\theta = 2^{\prime}$ Gaussian smoothing scale, using the galaxies in the third tomographic redshift bin. Top: The calculations of HSC-Y1 $\kappa$-maps (red dots), and predictions from Minkowski functionals assuming a Gaussian random field (GRF; dashed line). \newtext{Bottom: The difference between the data measurements and Gaussian predictions $V_k - V_k^{\rm G}$ divided by the maximum of the Gaussian prediction $V_k^{\rm G,\,max}$ to help the visualization of the non-Gaussian features}. The uncertainties for our $V_0$, $V_1$, and $V_2$ measurements (red error bars) are calculated using the covariance matrix from our simulations.} \label{fig:MFs}

\end{figure*}

The large-scale structure of the Universe is the result of the evolution of matter fluctuations, which is well described by the $\Lambda$CDM model. At late times, the underlying matter density field creates large gravitational potentials that can be probed by the bending of light from background galaxies. The primary manifestation of this bending or lensing effect is referred to as cosmic shear (see \cite{Kilbinger_2015} and \cite{Mandelbaum_2018} for recent reviews), which is understood as the distortion and magnification of the observed galaxy shapes. The Stage-III\footnote{Stage-III definition introduced by the Dark Energy Task Force report \citep{albrecht2006report}.} galaxy surveys observing millions of extragalactic objects, such as Kilo-Degree Survey \citep[KiDS;][]{Hildebrandt_2020,Giblin_2021}, the Dark Energy Survey \citep[DES;][]{DES_2016,Troxel_2018,Amon_2022}, and the Hyper-Suprime Camera survey \citep[HSC;][]{Hikage_2019,Li_2022}, are used to constrain the Universe at late times. However, the constrained $S_8$ parameter, defined as $S_8 \equiv \sigma_8 (\Omega_{\rm m}/0.3)^{0.5}$, with $\Omega_{\rm m}$ the matter energy density, and $\sigma_8$ the fictitious standard deviation of matter fluctuations in spheres of $8h^{-1}\, \text{Mpc}$ if they had evolved linearly, is in small discrepancy. This disagreement arises when comparing galaxy lensing measurements with the value inferred from observations of the primordial Cosmic Microwave Background (CMB) \citep{Planck_2018,Bianchini_2020} as well as quasi-linear matter clustering (CMB lensing) \citep{Madhavacheril_2020}.

Over the last decade, various explanations have emerged to address the $S_8$ discrepancy, including important systematic effects in galaxy surveys, such as photometric redshift estimations \citep{Hidelbrandt_2017} and shear calibrations \citep{Melchior_2010}. Nevertheless, their impact has been studied comprehensively, and it is unlikely to account for the total $S_8$ difference \citep{Hildebrandt_2021,Mandelbaum_2018_b}. The source of this discrepancy still remains an open question, which could reveal unaccounted systematic effects or the need of complementary modeling to include smaller scales for the current measurements.

The statistical properties of the matter density field at late times are non-Gaussian, which makes the study of methods beyond two-point statistics indispensable for capturing the information from non-linear scales. Several measurements probing the shear and convergence field have been proposed in the last decade, including the bispectrum and three-point function \citep{Takada_Jain_2003,Fu_2014,Halder_2023}, the Probability Distribution Function (PDF) \citep{Liu_2019,Boyle_2021,Giblin_2021,Thiele_2023,anbajagane2023beyond}, measurements of peaks and minima \citep{Davies_2022,Zucher_2022,Liu_2023,Marques_2024}, density split statistic \citep{Gruen2018,Burger2023,Paillas2024}, Minkowski functionals \citep{Marques_2019}, Betti numbers \citep{Feldbrugge_2019}, and scattering transforms \citep{Cheng_2024,Valogiannis_2024,gatti2024dark}. Also, methods employing deep learning such as convolutional neural networks \citep{Fluri_2018,Gupta_2018,Fluri_2022,Zhong_2024} are equally competitive to recover the non-Gaussian information stored in the non-linear matter field. In this paper, we evaluate the application of Minkowski functionals using data from the HSC year 1 dataset.

 
Minkowski functionals provide an algebraic description of the geometrical properties of a field \citep{Mecke_1994}. In cosmology, they measure the features of the patterns formed by large-scale structure of the Universe \citep{Schmalzing_1996,Schmalzing_1997}. Such patterns trace the distribution of matter of the Universe, being sensitive to the higher-order moments of the matter density field \citep{Sato_2001}. Recently, Minkowski functionals have been used for studies of non-linear features in the CMB \citep{Lim_2012,Novaes_2016,CarronDuque_2024,Hamann_2024}, anisotropies in the distribution of galaxies \citep{2003PASJ...55..911H, Appleby_2022}, constraints on neutrino masses \citep{Marques_2019,Liu_2023_MF}, studies of HI gas distribution using 21-cm intensity mapping \citep{Spina_2021,Schimd_2024}, and modified gravity \citep{2017MNRAS.466.2402S, Jiang_2024}. In addition, Minkowski functionals are known as a novel and efficient probe for primordial non-Gaussianity as showed in \citet{2006ApJ...653...11H, 2008MNRAS.385.1613H, 2008MNRAS.389.1439H, 2012MNRAS.425.2187H, 2012ApJ...760...45S}.
For weak lensing studies, Minkowski functionals have been proposed to break the degeneracy between $\Omega_{\rm m}$ and $\sigma_8$ inherent in two-point statistics \citep{Matsubara_2001}. Moreover, more studies using Minkowski functionals on weak lensing fields provide insight on constraining cosmology beyond power spectrum \citep{Kratochvil_2012,Shirasaki_2014,Petri_2015,Grewal_2022}, including measurements using the Canada-France-Hawaii Lensing Survey \citep[CFHTLenS;][]{Heymans_2012} data. However, the analysis from \cite{Petri_2015} finds important biases in the $(\Omega_{\rm m},\, \sigma_8)$ parameter-space. In this paper, we present cosmological constraints derived from Minkowski functionals to evaluate the power of this statistic using the first-year HSC weak lensing data, and to confirm whether the trends observed in these previous studies.

This paper is organized as follows: In Section \ref{sec:MFs} we describe the application of Minkowski functionals to weak lensing maps. The used data of the HSC-Y1 survey is explained in Section \ref{sec:data}. We describe the simulations for making our predictions in Section \ref{sec:simulations}. \newtext{We provide a description of our modeling of Minkowski functionals using simulations in Section \ref{sec:methods}. We comment in the analysis of the studied systematics in section \ref{sec:systematics}}. The results are presented in Section \ref{sec:results}. Finally we summarize the results and draw conclusions in Section \ref{sec:conclusions}.

\section{Minkowski functionals}\label{sec:MFs}

Minkowski Functionals correspond to $d+1$ independent and unique functions, with $d$ the number of dimensions, that describe the geometrical properties of a mathematical space. These functions or ``gauges'' \newtext{measure} the size, shape and connectivity \newtext{of a manifold}. In the specific case of convergence $\kappa$-maps, which is a two-dimensional pixelated space, $d = 2$, these properties are the Area $V_0$, Perimeter $V_1$ and Genus $V_2$, which combines the Euler characteristic of the connected pixels or ``Island'', and the number of holes in the same space. These are measured in the normalized area $A$ (in our case, the sum of the area of all pixels) and can be defined as:

\begin{eqnarray}
    V_0 = \frac{1}{A} \int_{\Sigma(\nu)} da, \\
    V_1 = \frac{1}{4A} \int_{\partial \Sigma(\nu)} dl, \\
    V_2 = \frac{1}{2\pi A} \int_{\partial \Sigma(\nu)} \mathcal{K}dl,
\end{eqnarray}

where $da$ and $dl$ are the area and length elements, respectively, \newtext{$\mathcal{K}$ is the curvature of the boundary, and $\partial \Sigma(\nu)$ denotes the excursion set boundary}. For the $\kappa$-field we measured over the surface $\Sigma(\nu)$ with $\nu > \kappa/\sigma_0$, which is defined as a threshold for every value of $\kappa$, divided by the standard deviation $\sigma_0$ of simulated convergence maps in the fiducial cosmology. 

A few studies have described Minkowski functionals calculations for weak lensing convergence maps. \cite{Kratochvil_2012} describe the computation of Minkowski functionals for 2D pixelated weak lensing fields as robust, containing valuable non-Gaussian information. Even though the continuum underlying matter density is unknown and the Minkowski functionals can be noisy for a large pixel array, the integral over the amount of information is stable and informative, resulting in a reliable estimator. Each Minkowski functional provides information that can be qualitatively described as explained in \cite{Kratochvil_2012}: $V_0$ measure the cumulative distribution function (CDF) of the histogram for the pixel values of the maps (sensitive to the distribution of peaks). $V_1$ includes a delta function over the convergence values, which measure individual shapes of objects given a value of $\kappa$. $V_2$ measures the same shapes and their connectivity. However, none of these functions consider the spatial distribution of the pixel values, which is encapsulated by the power spectrum, making the information complementary between the two types of statistics. In such setup, the combination of power spectrum and Minkowski functionals should provide tighter constrains for both $\sigma_8$ and $\Omega_{\rm m}$ parameters as shown in \cite{Kratochvil_2012,Petri_2013,Shirasaki_2014,Marques_2019,Grewal_2022}. 

The shapes of the Minkowski functionals as function of $\nu$ are shown in Figure \ref{fig:MFs}. We compare measurements of the Minkowski functionals from the HSC-Y1 data (red dots), and their analytical predictions obtained from Gaussian random fields \citep{Adler_1981, 1986PThPh..76..952T, Schmalzing_1997} (black dashed lines), using 18 equally spaced bins in the range $-4 < \nu < 4$. For this comparison, we calculated the \newtext{difference between the Minkowski functionals calculated from the data, and their predictions assuming a Gaussian random field $V_k - V_k^{\rm G}$ (bottom panels), with $i=0,1,2$. This difference reveals the non-Gausian information nature of the data and can be predicted for the perturbative non-Gaussian case \citep{Matsubara_2003}. Most differences can be found when comparing $V_0$, noticing features deviating from the Gaussian predictions for $\nu > -2$, being largely deviated for the high values of $\nu=0$. For our $\kappa$ measurements this difference seems to follow a simpler profile proportional to $ e^{-\nu^2 / 2}$ (dotted line). In general, the features showed by $V_0$, $V_1$ and $V_2$ are consistent with the ones found in \cite{Petri_2013}, showing that the difference $V_i - V_i^{\rm G}$ does not converge to the perturbative non-Gaussian case for small smoothing scales, like the ones used throughout this study.} 

\section{HSC Year one weak lensing data}\label{sec:data}

We use the weak lensing shear catalogues from HSC-Y1 \citep{Mandelbaum_2017} with redshift estimated from the HSC five broad-band photometry using the {\sc mlz} code \citep{Tanaka_2018}. Then we applied tomographic redshift cuts in 4 bins with edges [0.3,0.6,0.9,1.2,1.5], but neglecting the fourth redshift bins due to an unknown systematic error found prior to un-blinding. For example, the impact of including this tomographic redshift bin is shown in \citep{Thiele_2023}, where a more significant shift on higher $S_8$ is found for the PDF analysis. Gaussian smoothing is performed over the shear maps with different smoothing scales, using $\theta_{\rm s} = {1,2,4,5,8,10}$ arcmin ($^\prime$). Such scales are chosen to study the non-Gaussian features of the $\kappa$-maps, while mitigating possible systematic effects at small scales, such as baryonic physics. Finally, we go from tangential shear to the flat-sky convergence $\kappa$ maps using the Kaiser-Squires inversion method \citep[KS;][]{Kaiser_Squires}, after applying map in-painting on the masked pixels \citep{Pires_2009,Starck_2021} to reduce undesired artifacts created by the survey mask. A detailed description of the convergence field reconstruction can be found in \cite{Marques_2024}. 

\begin{figure}
    \centering
    \includegraphics[width=\linewidth]{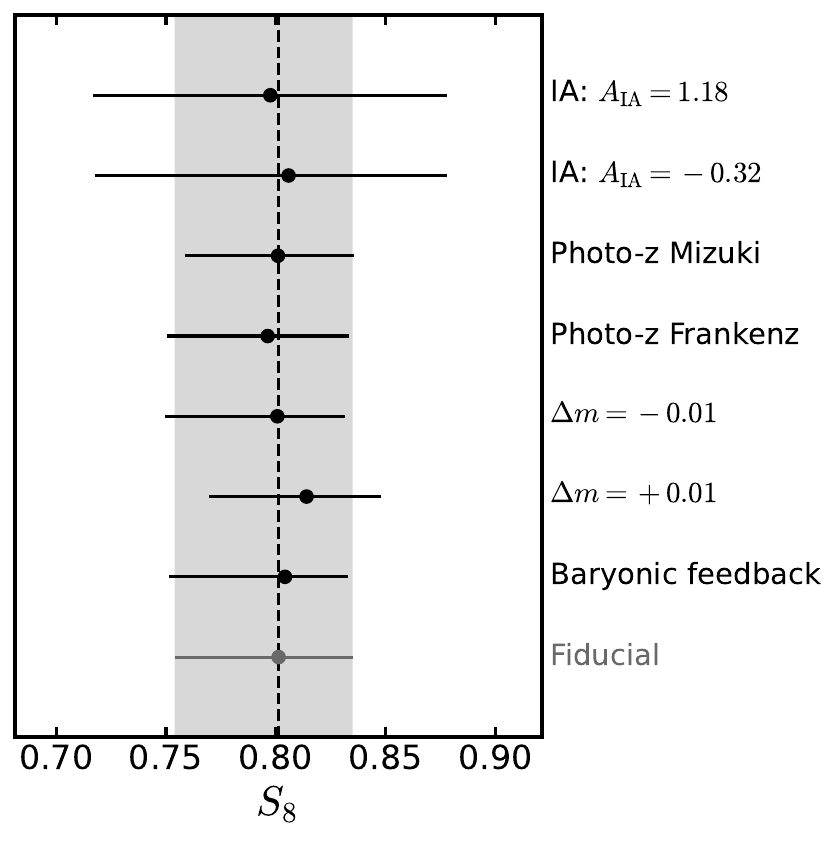}
    \caption{Analysis of the impact of several systematic effects in the inference of the $S_8$ parameter. Considering the fiducial value (dashed line), we contaminate the data vector with a systematic as labeled in the text. Then we use this new `observation' (black dot with error bar) to repeat the inference in these new values for $S_8$. We add the recovered fiducial inference data point (grey dot) and the 68\% confidence interval (shaded area) to visually help the comparison. 
    }
    \label{fig:blind_systematics}
\end{figure}

\section{Simulations}\label{sec:simulations}

To study the non-Gaussian signal provided by the non-linear evolution of the field we use numerical simulations. These N-body simulations are divided into two sets: One set is used to calculate the covariance matrix and test different systematic effects in our measurements, which is defined  as the fiducial cosmology. The second, is a suite of simulations with 100 different cosmologies for emulation.

For the calculations of the covariance matrix, simulations come from a set of 108 quasi independent full-sky simulations \citep{Takahashi_2017}, which are utilized to make 2268 mock map realizations of the HSC-Y1 shape catalogues with a fiducial cosmology ($\Omega_{\rm m}= 0.297$ and $\sigma_8 = 0.82$) obtained from the Wilkinson Microwave Anisotropy Probe nine-year data \citep{Hinshaw_2013}. Mock realizations account for many survey properties (including the survey mask) following the methodology of \cite{Shirasaki_2019}. 

To create the Minkowski functionals predictions from different cosmology values, we use simulations from \cite{Shirasaki_2021}, which provide ray-traced maps that are used to create mock catalogues based on these simulations with 100 pairs of $\Omega_{\rm m}$ and $S_8$ values (the remaining cosmological parameters are fixed from the fiducial cosmology). Each cosmology-varied simulation includes different observers in the periodic box to obtain 50 quasi-independent realizations. The same procedure to include observational aspects to the fiducial set is implemented to obtain the final cosmology-varied simulations. The full methodology to create the mock catalogues and convergence maps from the simulations is explained in \cite{Marques_2024} and it has been used in previous studies of non-Gaussian statistics \citep{Thiele_2023,Cheng_2024,Grandon_2024,Novaes_2024}.

\section{Minkowski functionals application to HSC-Y1 data} \label{sec:methods}
In this section, we overview the methodology to calculate Minkowski functionals in the HSC-Y1 data, including the model predictions, the covariance matrix and the defined likelihood for the inferred posterior. As mentioned in \cite{Marques_2024}, we follow a blind procedure, using simulations to test the pipeline and define our baseline analysis by selecting smoothing lengths and scale cuts on the data vector. 

\subsection{Emulator for Minkowski functionals}

\begin{figure}
    \centering
    \includegraphics[width=0.8\linewidth]{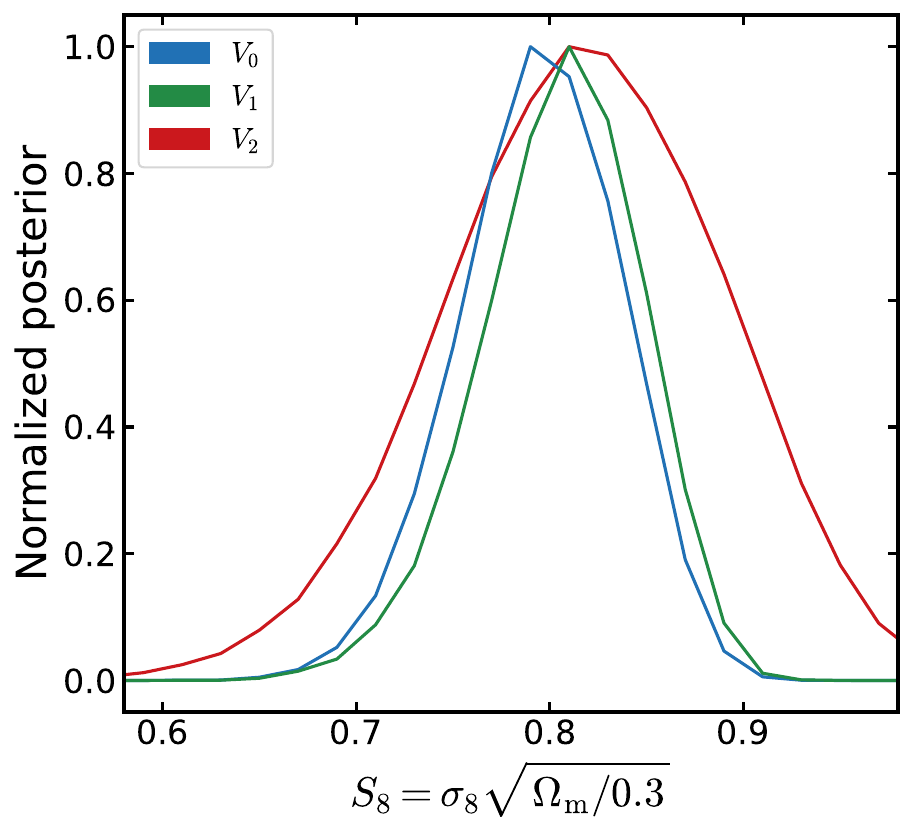}\\
    \includegraphics[width=0.8\linewidth]{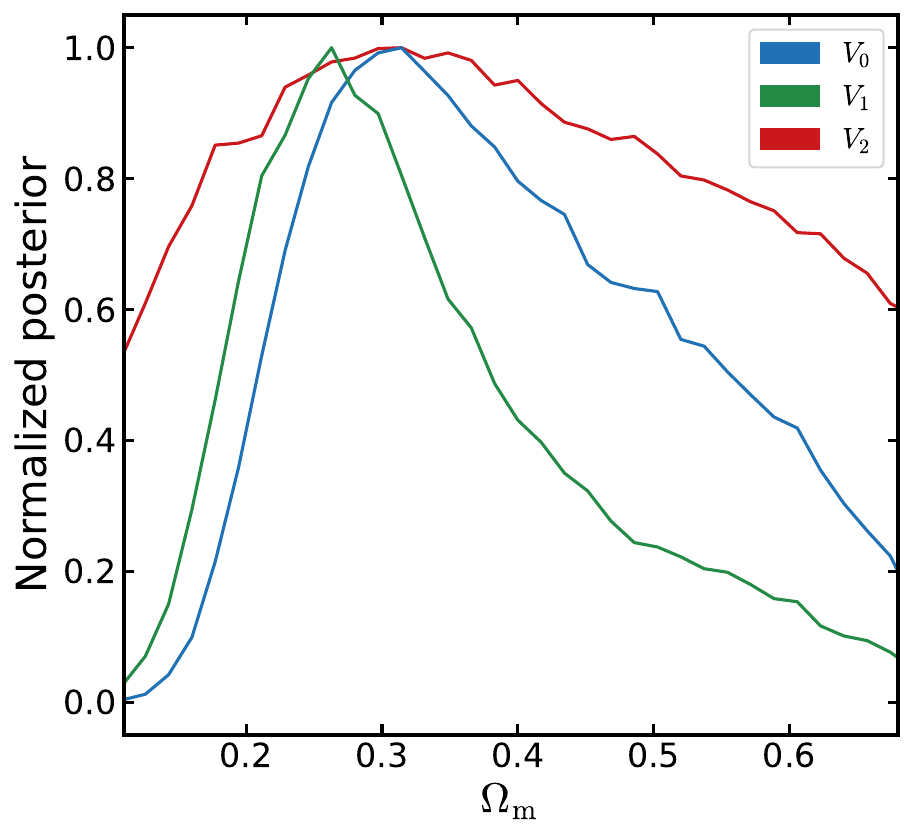}
    \caption{Individual posteriors of $S_8$ \newtext{(top) and $\Omega_{\rm m}$ (bottom)} parameters using the individual Minkowski functionals: $V_0$ (blue line), $V_1$ (green line) and $V_2$ (red line). Most of the $S_8$ information is contained in $V_0$ (area) and $V_1$ (perimeter) \newtext{with a improved constraints on $\Omega_{\rm m}$ for $V_1$}.}
    \label{fig:MFs_V0_V1_V2_comparison}
\end{figure}

An emulator for our summary statistic is designed from the cosmology-varied simulations using Gaussian process regression (GPR). Using a slightly modified function than the one used in \cite{Marques_2024}, with a different radial basis kernel value, we emulate the data vector using the mean of the 50 realizations for each cosmology (out of 100 cosmology pairs). Then we are able to produce emulated Minkowski functionals and the angular power spectrum data vector from a given cosmology pair with a determined accuracy. To validate the emulator, we use the ``leave-one-out" method, where the emulator is trained using 99 cosmology pairs, predicting the one not used for training. After testing, the emulator can obtain accurate (between $2-3\%$ of differences) model predictions for the Minkowski functionals in an unbiased way. The accuracy decreases for the bins on the extremes of the data vector (low and high tail values of kappa), which suffer more from statistical errors like sample variance and shot noise, but not higher than 5\% in average.

\subsection{Likelihood estimation}

Considering the large number of bins of the combined statistics, which includes the power spectrum and Minkowski functionals for the three tomographic bins and several smoothing scales, we use {\sc moped} compression \citep{Heavens_2000,Heavens_2017} and Gaussian likelihood approximation for the inference analysis. First, we define the compressed data vector, which approximately preserves information from the original Minkowski functional data vectors, while reducing its dimensionality to the number of free parameters we are trying to infer ($\Omega_{\rm m}$ and $S_8$). By compressing the data vector, the likelihood is close to a Gaussian distribution, due to central limit theorem. In addition, the power spectrum in our analysis is measured for a multipole scale of $300 < \ell < 1000$, in four data bins, also with a likelihood which is well approximated by a Gaussian distribution. Then the combined likelihood is then defined as:
\begin{equation}
    -2 \log \mathcal{L} = (\mathbf{x} - \bm{\mu(\theta)})^{\intercal} \mathbf{\Sigma^{-1}} (\mathbf{x} - \bm{\mu(\theta)})\,,
\end{equation}
where $\mathbf{x}$ is the vector measured on the data, $\bm{\mu(\theta)}$ is the prediction from the emulator for the cosmological parameter vector $\bm{\theta}$, and $\mathbf{\Sigma^{-1}}$ is the inverted covariance matrix from the fiducial simulations. For the data vector $\mathbf{x}$ we use the weighted averaged on all the HSC-Y1 fields, as described in \cite{Marques_2024}.

The sampling is performed using the publicly available code {\sc cobaya},\footnote{\url{https://github.com/CobayaSampler}} which creates Monte Carlo Markov Chain samples using the Metropolis-Hasting algorithm. We use uniformly distributed priors for the cosmological parameters, $0.11 < \Omega_{\rm m} < 0.65$ and $0.5 < S_8 < 1.1$, which is fairly covered by the cosmology-varied simulations.

\section{Systematic effects}\label{sec:systematics}

\begin{figure}
    \centering
    \includegraphics[width=\columnwidth]{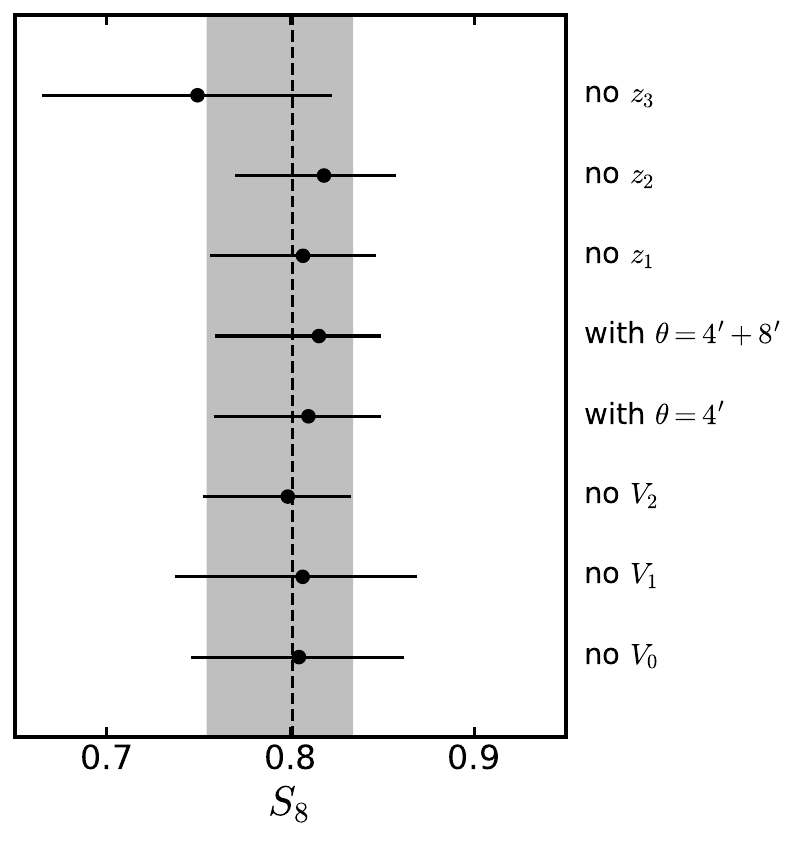}
    \caption{Consistency check testing the different choices of the used data vectors for calculating the posterior of $S_8$ (black points and lines) contrasted with the baseline results (dashed line) including the 68\% confidence interval (grey shaded area) .}
    \label{fig:cosisntency_checks}
\end{figure}

\begin{figure*}
    \centering
    \includegraphics[width=\columnwidth]{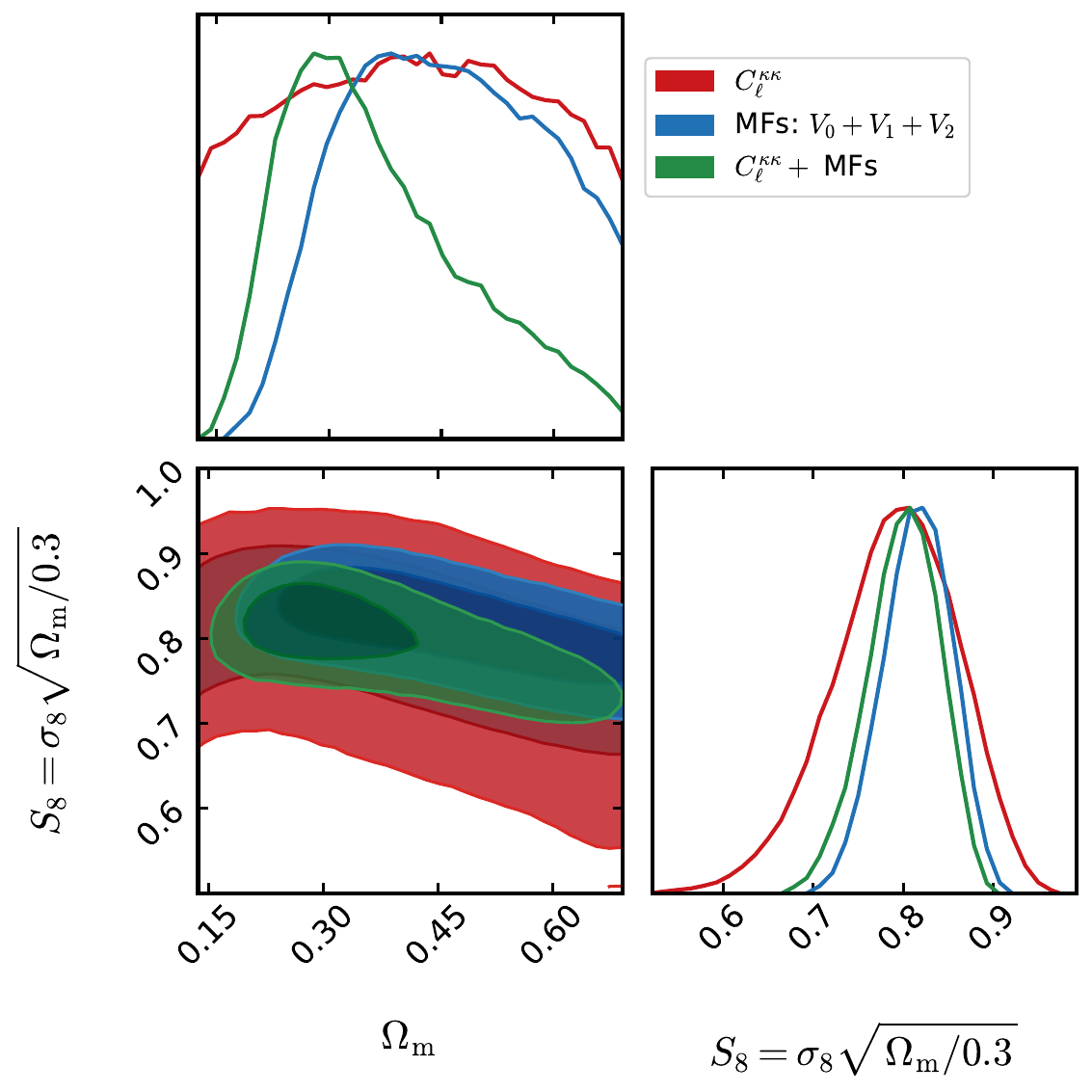}
    \caption{Constraints of Minkowski functionals for $\Omega_{\rm m}$ and $S_8 = \sigma_8 \sqrt{\Omega_{\rm m} / 0.3}$ parameters using the HSC-Y1 data. The inferred posterior is calculated using the angular power spectrum $C_{\ell}^{\kappa\kappa}$ (red), the three Minkowski functionals (blue) and the joint measurement using both statistics (green). We colour the 68\% (inner line) and 95\% (outter line) confidence intervals contours.}
    \label{fig:Cell+MFs_cp}
\end{figure*}

We test the impact of systematic effects that are not included in the modelling and can potentially bias the final result of our cosmological inference pipeline. These physical effects can be added as contamination directly on the data vector. To take into account errors in calibration, we add a shifting of $\pm 1\%$ in the multiplicative bias factor, which correspond to the uncertainty measured from simulations \citep{Mandelbaum_2018_b}. We also include different photometric redshift estimation codes ({\sc frankenz} and {\sc mizuki}) in our pipeline. To study the possible shift on $S_8$ due baryonic effects, we calculate the Minkowski functionals on the $\kappa$TNG dataset \citep{Osato2021}, which consists of a set of 10,000 convergence maps using dark matter only (DMO) and hydrodynamic simulations from the IllustrisTNG model \citep{Nelson_2019}. Then, we multiply our data vector by the ratio of our calculations, which tells what scales in our statistic are more affected by the effect of baryonic physics. Similarly, the effect of intrinsic alignments (IA) is added from mock shape catalogues infused with the non-linear tidal alignment model \citep[NLA;][]{Bridle_2007}, with $A_{\rm IA} = 1.18$ and $A_{\rm IA} = -0.32$ \citep{Harnois-Deraps2021}, where we compare the statistics for the mocks with and without the effect of IA, adding the ratio to the fiducial data vector. We do this comparison for results of the $\theta=2\prime$ Gaussian kernel smoothed map, which is our baseline analysis, and provides enough information of non-Gaussian statistics.  The results of the individual impact for each systematic applied to the fiducial cosmology statistic are shown in Figure \ref{fig:blind_systematics}. We find no significant deviation from the fiducial value of $S_8$ for the studied systematic effects, finding values between $0.05$-$0.38\sigma$ for such shift. However, the initial inference analysis revealed a systematic effect for the individual Minkowski functional $V_0$ at the first redshift bin $z_1$, showing significantly wider contours, which are preserved when combined with $V_1$ and $V_2$. To solve this particular issue, we keep the last 4 $\nu$ bins of this data vector. We select this cut after checking that $V_0$ at $z_1$ has negligible constraining power and the posteriors of our final blind analysis is unchanged. We decide by consistently applying scale cuts on the tails of $V_0$, $V_1$ and $V_2$ in all redshift bins, removing the first and last bins of each individual data vector (with the exception of $V_0$ at $z_1$).

\section{Results}\label{sec:results}

We check for the individual constraining power of $V_0$, $V_1$ and $V_2$, and different data vector combinations to test the robustness of our constraints. Then, we present the unblinded results after applying the scale cuts to avoid the impact of the systematic effects, showing the final constraints for angular power spectrum, Minkowski functionals and the combined statistics.

\subsection{Individual Minkowski functional contribution}

We provide constraints of individual Minkowski functionals for the fiducial cosmology as a target, trying to discern if there is any specific function that outperforms the others, instead of combining them. We show their comparison in Figure \ref{fig:MFs_V0_V1_V2_comparison}, where the 1D marginal posterior of $S_8$ \newtext{(top) and $\Omega_{\rm m}$ (bottom) are} plotted for the individual Minkowski functionals: $V_0$ (blue line), $V_1$ (green line) and $V_2$ (red line). \newtext{These show a better performance of $V_1$ for constraining $\Omega_{\rm m}$, but} individually similar results for both the area and perimeter, \newtext{and} wider constraint provided by the genus function $V_2$, meaning that $V_0$ and $V_1$ contain more non-Gaussian information than $V_2$. Additionally, we notice the constraining power of $V_2$ is at the level of power spectrum, which is opposite to what is found in \citep{Kratochvil_2012}, where genus has similar constraining power compared with $V_0$ and $V_1$. However, their analysis is bases on Fisher forecast, where the statistical errors are neglected and the contours can be overconfident. In the same line, $V_0$ is expected to contain information from the 1-point function measured by the PDF of $\kappa$. Indeed, by comparing the $1\sigma$ values from the $V_0$, we find $S_8 = 0.799_{-0.045}^{+0.036}$ consistent with some of the results presented on \cite{Thiele_2023}, specifically when PDF is combined with two-point statistics. We also find $S_8 = 0.816_{-0.051}^{+0.029}$ for $V_1$ and $S_8 = 0.818_{-0.070}^{+0.069}$ for $V_2$.

\subsection{Consistency checks}

We investigate how small variations in the choice of the data vector impact our cosmological constraints. For this, we implement some consistency checks performed by previous analyses applying non-Gaussian statistics to HSC data \citep{Thiele_2023,Marques_2024,Cheng_2024,Novaes_2024}. We provide a simple test of consistency in Figure \ref{fig:cosisntency_checks}, where from the baseline analysis using $\theta = 2^\prime$, we remove individual Minkowski functionals, add an individual smoothing scales of $\theta = 4^\prime$ and $\theta = 8^\prime$, and remove each of the tomographic redshift bins. By comparing with the obtained baseline constraint (dashed line and grey area), we list the different choices of the constraints obtained by the new data vectors. No substantial shift (higher than $1\sigma$) from the baseline result is found if we consider the uncertainties of each data point, even though there is an important shift towards smaller $S_8$ value once the third redshift bin ($z_3$) is removed from the analysis. These results show consistency with the analysis from \cite{Marques_2024} and \cite{Thiele_2023}, where the same tendency is found using different statistic, and from \cite{Novaes_2024} which also uses Minkowski functionals data. Additionally, we find that the data vector including larger smoothing scales, obtains slightly higher $S_8$ values, but with a larger uncertainty. Opposite tendency is found in \cite{Kratochvil_2012} (again, using Fisher forecast), where the combination of several smoothing scales improve the constraints. 

Finally, removing individual Minkowski functionals complements the results of Figure \ref{fig:MFs_V0_V1_V2_comparison}, where we found that $V_0$ and $V_1$ have more constraining power than $V_2$. These results reaffirm that slightly tighter constraints are provided by $V_0 + V_1$, and $V_2$ have the provides less information. We also report marginally wider constraints when removing $V_1$ from the data vector.

\subsection{Angular power spectrum and Minkowski functional constraints} \label{sec:results:constraints}

Figure \ref{fig:Cell+MFs_cp} shows the constraints for $S_8$ and $\Omega_{\rm m}$ using the HSC-Y1 data. We compare the posterior of these parameters calculated using both the angular power spectrum $C_{\ell}^{\kappa\kappa}$ (red contours), the Minkowski functionals (blue contours), and the combination of these statistics (green contours).  For the two-point statistics only, $C_{\ell}^{\kappa\kappa}$ constraints for $\Omega_{\rm m}$ are prior dominated, whereas $S_8 = {0.790}_{-0.057}^{+0.077}$, which is consistent with previous results \citep[e.g.][]{Hikage_2019}. Including Minkowski functionals in the analysis results in tighter constraints, with a significant improvement on $S_8$. The marginal posterior of the combined statistics yields $\Omega_{\rm m} = {0.293}_{-0.043}^{+0.157}$ and $S_8 = {0.808}_{-0.046}^{+0.033}$, roughly a $40\%$ tighter than using the power spectrum only for $S_8$. It is worth noticing that using only the Minkowski functional for the inference, results in virtually the same constraints improvements for $S_8 = {0.819}_{-0.048}^{+0.032}$. From these results, we also find noticeable constraints on $\Omega_{\rm m}$ from the combination of our statistics, indicating the power of Minkowski functionals to break degeneracy between $S_8$ and $\Omega_{\rm m}$, which is not possible from two-point statistics only. These results show agreement with different non-Gaussian probes on the same dataset.

\section{Summary \& Conclusions}\label{sec:conclusions}

We apply Minkowski functionals, a type of non-Gaussian statistic, to convergence maps built from the HSC-Y1 dataset, obtaining new constraints on $S_8$ and $\Omega_{\rm m}$. We combine these measurements with two-point statistics, utilizing measurements of the angular power spectrum of convergence $C_{\ell}^{\kappa\kappa}$ to improve the results from two-point functions only.
Using a likelihood approach based on emulated simulation predictions, with different $S_8$ and $\Omega_{\rm m}$, we do a full posterior analysis to find constraints on these cosmological parameters. Before un-blinding the data, we calibrate scale cuts on the individual data vectors and Gaussian smoothing scales of the maps to mitigate biases coming from systematic effects that can affect the final results once we include small scale modes. However, no significant shift is found for the studied systematics, which indicates these are effectively controlled by our scale cut choices. This is in agreement with results from \cite{Grandon_2024}, which investigates the effect of baryon in several non-Gaussian statistics for HSC-Y1 dataset in the same regime used for this study. Additionally, we test the robustness of our analysis by checking the self-consistency of this probe, by measuring different choices of the data vector, including constraints using individual Minkowski functionals, adding additional smoothing scales, removing part of the data vector, and removing individual tomographic redshift bins. The baseline analysis includes $V_0$, $V_1$ and $V_2$, combined with measurements of $C_{\ell}$ at $300 < \ell < 1000$, with a smoothing scale of $\theta = 2^\prime$, yielding $S_8 = {0.808}_{-0.046}^{+0.033}$, consistent with other probes of non-Gaussian statistics using HSC-Y1 data and with the $S_8$ inferred by the CMB probes. This value for $S_8$ represents a $40\%$ improvement in comparison to the constraint provided by power spectrum only. Also, the constraint $\Omega_{\rm m} = {0.293}_{-0.043}^{+0.157}$ is found with similar results to the ones presented in \cite{Marques_2024} and \cite{Cheng_2024}, using peak statistics and scattering transform, showing the improved performance of these statistics to break degeneracy between $\Omega_{\rm m}$ and $\sigma_8$. Also, the final constraints are fully consistent with \cite{Novaes_2024}, which uses a different compression method for the data vector and a machine-learning based method for the posterior estimation.

\newtext{In contrast to the power spectrum  that can fully characterize a random Gaussian fields, Minkowski functionals are a way to encapsulate the information from non-Gaussianity terms \citep{Matsubara_2003}}. We find that most of the constraining power on \newtext{$\Omega_{\rm m}$ and} $S_8$ from our combined statistic, comes from the area $V_0$ and perimeter $V_1$, each with very similar constraining power, \newtext{but a with better constraint of $\Omega_{\rm m}$ given by $V_1$}. Moreover, for the genus $V_2$ these constraints are wider, which indicates this particular Minkowski functional capturing less non-Gaussian information from the HSC dataset. We believe that statistical errors and data noise are relevant and have a higher impact on $V_2$, \newtext{in particular on the second and third derivatives of the skewness parameter $S_1$ and $S_2$, which contribute to the non-Gaussian information of $V_2$ \citep{Petri_2013}, indicating a larger impact to of noise, to the one quantified in \cite{Matsubara_2001}}. In fact, \cite{Shirasaki_2013} shows that even though is a robust measurement, either cosmic variance or statistical error may affect on the individual Minkowski functionals. Additionally, a similar tendency is found in \cite{Petri_2013,Petri_2015}, with $V_2$ providing less constraining power in comparison to $V_0$ and $V_1$, \newtext{and showing a slow convergence of the perturbative series for small smoothing scales ($\theta < 15^{\prime}$), such as the ones used in this study}. Regardless, removing $V_2$ from the analysis does not improve the constraints. Additionally, the analysis shows that the most constraining power comes from the third redshift bin $z_3$ of the data, because of the larger sample this particular redshift bin provides. The same tendency found in the previous studies using the HSC-Y1 data.

In general, these results improve the previous constraints from \cite{Petri_2015} using the CFHTLens dataset, where a similar approach using emulation is used to model the cosmological dependence of Minkoswki functionals. Our study finds no biases in the constraints results for the ($\Omega_{\rm m}$, $\sigma_8$). However, their analysis is also extended to more cosmological parameters, such as the dark energy equation of state $w$, which can drive the bias found in their study. 

Finally, we believe that future analyses using Stage-IV galaxy surveys, such as Vera Rubin Observatory Legacy Survey of Space and Time \citep[LSST;][]{Ivezi_2018}, Euclid \citep{Laureijs_2011}, and Nancy Grace Roman Space Telescope \citep{Eifler_2021}, with a better control of the photometric redshifts and larger surveyed area, will provide a more accurate result.

\section*{Acknowledgements}
JA thanks useful discussion with Surhud More, Linda Blot, Joachim Harnois-Déraps and Masahiro Takada. JA is supported
by JSPS KAKENHI Grant Number JP23K19064.
MS is in part supported by MEXT KAKENHI Grant Number (20H05861, 23K19070, 24H00215, 24H00221).
CPN thanks Instituto Serrapilheira for financial support.
Numerical computations were in part carried out on Cray XC50 at Center for Computational Astrophysics, National Astronomical Observatory of Japan. This work was supported by JSPS KAKENHI Grants 23K13095 and 23H00107 (to JL). 
This research used computing resources at Kavli IPMU. This research used resources at the National Energy Research Scientific Computing Center (NERSC), a U.S. Department of Energy Office of Science User Facility located at Lawrence Berkeley National Laboratory, operated under Contract No. DE-AC02-05CH11231. 
The Kavli IPMU is supported by the WPI (World Premier International Research Center) Initiative of the MEXT (Japanese Ministry of Education, Culture, Sports, Science and Technology). This manuscript has been authored by Fermi Research Alliance, LLC under Contract No. DE-AC02-07CH11359 with the U.S. Department of Energy, Office of Science, Office of High Energy Physics.

\section*{Data Availability}

 The HSC-Y1 data is publicly available. The simulations and the generated mock catalogues used in this paper are available on reasonable request. 



\bibliographystyle{mnras}
\bibliography{references} 




\appendix

\bsp	
\label{lastpage}
\end{document}